\documentstyle[manuscript,aps,epsf]{revtex}
\begin{document}
\preprint{UCONN-97/12}

\title{Supersymmetry Breaking with Periodic Potentials}

\author{Gerald Dunne\footnote{dunne@hep.phys.uconn.edu}}
\address{Physics Department, University of Connecticut, Storrs, CT 06269}
\author{Jake Mannix\footnote{jake@dirac.ucsc.edu}\footnote{Supported through the National Science Foundation R.E.U. (Research Experience for Undergraduates) program at the University of Connecticut.}}
\address{Physics Department, University of California, Santa Cruz, CA 93106}
\maketitle

\begin{abstract}
We discuss supersymmetry breaking in some supersymmetric quantum mechanical models with periodic potentials. The sensitivity to the parameters appearing in the superpotential is more acute than in conventional nonperiodic models. We present some simple elliptic models to illustrate these points.
\end{abstract}

\vskip 1cm

Supersymmetry (SUSY) breaking is an outstanding problem that is directly relevant for the application of SUSY to particle physics \cite{witten}. Certain aspects of this symmetry breaking mechanism may be studied using the simplest SUSY model, that of SUSY quantum mechanics \cite{witten}. Typically, one considers models with discrete spectra \cite{witten}, while the extension to spectra with both bound and continuum states is relatively straightforward \cite{akhoury}. Recently, models with periodic potentials, which therefore have {\it band} spectra, have been considered \cite{dunne}. The main new feature is that it is possible for the periodic isospectral bosonic and fermionic potentials to have {\it exactly} the same spectrum, {\it including zero modes}. Thus it is possible to have models with unbroken SUSY but for which the bosonic and fermionic hamiltonians have exactly identical spectra \cite{dunne}. This is in contrast to the usual (nonperiodic and fast decaying) case for which at most one potential of an isospectral pair can have a zero mode. In this paper, we consider the breaking of SUSY in models with periodic superpotentials, and the sensitivity of these models to the parameters appearing in the superpotential. 

SUSY quantum mechanics on the real line can be summarized as follows \cite{witten}. The bosonic and fermionic Hamiltonians $H_\pm$ correspond to an isospectral pair of potentials $V_\pm(x)$ defined in terms of the ``superpotential'' $W(x)$ as
\begin{equation}
V_\pm(x)=W^2(x)\pm W^\prime(x)
\label{iso}
\end{equation}
The Hamiltonians may be factorized into products of hermitean conjugate operators as
\begin{equation}
H_+ = [{d\over dx} + W(x)][-{d\over dx}+W(x)]\,,\quad H_- = [-{d\over dx} + W(x)][{d\over dx}+W(x)]
\label{factorized}
\end{equation}
which indicates that $H_\pm$ are formally positive operators. The factorization
(\ref{factorized}) also implies that $V_\pm$ have (almost) the same spectrum because there is a one-to-one mapping between the energy eigenstates $\psi_E^{(\pm)}$:
\begin{eqnarray}
\psi_E^{(+)}={1\over \sqrt{E}}\left({d\over dx}+W(x)\right) \psi_E^{(-)};\qquad
\psi_E^{(-)}={1\over \sqrt{E}}\left(-{d\over dx}+W(x)\right) \psi_E^{(+)}
\label{mapping}
\end{eqnarray}
This mapping between states does 
not apply to the ``zero modes'' (eigenstates with $E=0$), which due to the 
positivity of $H_\pm$, are the lowest possible states in the spectrum. From (\ref{factorized}) it is easy to see that the Schr\"odinger equation $[-\partial_x^2 +V_\pm(x)]\psi_E^{(\pm)}=E\psi_E^{(\pm)}$ has zero modes 
\begin{equation}
\psi_0^{(\pm)}(x)=e^{\pm\int^x W}
\label{zero}
\end{equation}
provided these functions $\psi_0^{(\pm)}$ belong to the Hilbert space. SUSY is unbroken if at least one of the $\psi_0^{(\pm)}$ is a true zero mode. Otherwise, SUSY is said to be broken. In the broken SUSY case there are no zero modes and so the spectra of $V_\pm$ are identical [due to the mapping (\ref{mapping})].

The Witten index, $\Delta=tr(-1)^F=n_+-n_-$, counts the {\it difference} between the number $(n_+)$ of bosonic zero modes $\psi_0^{(+)}$ and the number $(n_-)$ of fermionic zero modes $\psi_0^{(-)}$ \cite{witten}. It is often a useful indicator of SUSY breaking: if $\Delta\neq 0$ then there must be at least one zero mode and so SUSY is unbroken; if $\Delta=0$ then more information is needed, as this could be either because there are no bosonic or fermionic zero modes at all (in which case SUSY is broken), or it could be because $n_+=n_-\neq 0$ (in which case SUSY is unbroken). Much of the power of the index method comes from the fact that the index may be reliably and easily computed in many cases, both in SUSY quantum mechanics and in SUSY field theory. To a large extent $\Delta$ is independent of the parameters (masses, couplings, volume, ...) of the theory; a property that allows the computation of $\Delta$ in special parametric regimes. Several direct applications of this are discussed in \cite{witten}.

This insensitivity to fine-tuning of parameters must, of course, be qualified. It is assumed that under the parametric changes the energy eigenstates move about continuously in energy - they do not suddenly appear in, or disappear from, the Hilbert space. The simplest example is to take the superpotential to be $W(x)=x-a$, where $a$ is some constant. This is just the harmonic oscillator system and it is clear that $H_-$ has a normalizable zero mode for any finite value of the parameter $a$, while $H_+$ has no normalizable zero mode for any value of $a$. So SUSY is unbroken for all values of the parameter $a$. On the other hand, with $W(x)=x^2-a$, it is clear that neither $H_+$ nor $H_-$ has a normalizable zero mode for any value of $a$ and so SUSY is broken for all $a$. With $a>0$ this is an example of {\it dynamical} SUSY breaking because the tree-level potential $W^2(x)$ does have zeros \cite{witten}. 

These simple examples with discrete spectrum may be generalized to include also continuum states. Consider 
\begin{equation}
W(x)=tanh\, x-a 
\label{tanhh}
\end{equation}
For $-1<a<1$, the fermionic hamiltonian $H_-$ has a normalizable zero mode $\psi_0^{(-)}(x)=e^{-a x}sech\, x$, while $H_+$ has no zero mode. Thus SUSY is unbroken. But for $|a|\geq 1$ the zero mode $\psi_0^{(-)}$ of $H_-$ becomes non-normalizable; there are therefore no zero modes, and SUSY is therefore broken. This discontinuous change comes about because for $|a|\geq 1$ the asymptotic limits of the superpotential have changed from having opposite signs to having the same sign. Correspondingly the behavior of $V_-$ at $|x|=\infty$ has been altered. Nevertheless, in this example there is still a finite range $(-1,1)$ of $a$ for which SUSY remains unbroken. On the other hand, consider the superpotential
\begin{equation}
W(x)=tanh^2\, x-a 
\label{tanhsq}
\end{equation}
Clearly there is no $a$ for which either $H_\pm$ has a zero mode, and so SUSY is broken for all values of the parameter $a$. We shall return to this example later.

For periodic potentials the situation is rather different. The criterion for SUSY breaking reduces, as before, to the question of whether the zero modes in (\ref{zero}) are elements of the Hilbert space. For nonperiodic systems on the real line this is a question of normalizability of these zero-mode wavefunctions. This can be phrased in terms of the asymptotic limits of $W(x)$, or in terms of the number of zeros of $W(x)$ being odd or even, or in terms of whether $W(x)$ is an odd or even function \cite{witten}. For periodic systems, normalizability is not the issue; rather, the zero-modes in (\ref{zero}) must be Bloch functions, and since they are zero energy they must in fact be {\it periodic} functions (with the same period as the potential) \cite{kittel}. It is easy to see from (\ref{zero}) that this translates into the requirement \cite{dunne} that the superpotential $W(x)$ satisfy:
\begin{equation}
\int_{\rm period}\, W(x) =0
\label{condition}
\end{equation}
 
This condition has two simple, but significant, consequences. First, suppose the superpotential $W$ is such that $\int_{\rm period}W\neq 0$. Then we can always arrange for $\int_{\rm period}W= 0$ simply by subtracting an appropriate finite constant from $W$. Conversely, suppose the condition (\ref{condition}) {\it is} satisfied; then we have no freedom to shift $W$ by any finite constant $a$, without breaking SUSY. 

We now illustrate these consequences with some examples that generalize those already mentioned for the nonperiodic case. First, consider the elliptic superpotential (see \cite{dunne})
\begin{equation}
W(x)=m \, sn(x|m)\, cd(x|m)-a
\label{ell-1}
\end{equation}
where $sn(x|m)$ and $cd(x|m)\equiv cn(x|m)/dn(x|m)$ are standard Jacobi elliptic functions \cite{abram}, and $m$ is the elliptic parameter ($0< m\leq 1$). The superpotential (\ref{ell-1}) has period $2 K(m)$, where $K(m)$ is the ``real elliptic quarter period''. Note that when $m=1$ the Jacobi functions reduce to hyperbolic functions: $sn(x|1)=tanh\,x$ and $cd(x|1)=1$. Thus, the superpotential in (\ref{ell-1}) reduces to $W=tanh\, x-a$ which is just the example discussed earlier in (\ref{tanhh}).

Now, since $\frac{d}{dx}[log\, dn(x|m)]= -m\, sn(x|m)\, cd(x|m)$, and $dn(x|m)$ has period $2K(m)$, we find that
\begin{equation}
\int_{\rm period} W = -2 a K(m)
\end{equation}
This only vanishes for $a=0$, in which case SUSY is unbroken \cite{dunne}. SUSY is broken for any nonzero value of the parameter $a$. Thus the periodic model in (\ref{ell-1}) is more sensitive to fine-tuning of the parameter $a$ than is the nonperiodic model with superpotential (\ref{tanhh}). This occurs even though there is no noticeable effect in the periodic system on the number of zeros of $W$, or on the values of $W(x)$ at the edges of a period, when $a$ deviates from $0$. 

Next, consider the periodic superpotential, with period $K(m)$,
\begin{equation}
W(x)=m^2 \, sn^2(x|m)\, cd^2(x|m)-a
\label{ell-2}
\end{equation}
When $m=1$, $W$ reduces to $tanh^2x-a$, which coincides with the earlier example (\ref{tanhsq}) for which SUSY was always broken, for all values of the parameter $a$. Here the situation is different - to determine whether SUSY is broken or not we look to the condition (\ref{condition}). 

Note the following facts (see \cite{abram}): (i) $\int^x W=(2-m-a)x+m\, sn(x|m)\, cd(x|m)- 2E(x|m)$, where $E(x|m)$ is the elliptic integral of the second kind: $E(x|m)\equiv\int_0^x dn^2(x|m)$ ; (ii) under a period shift, $E(x+K(m)|m)=E(x|m)-m\,sn(x|m)\,cd(x|m)+E(m)$, where $E(m)$ is the complete elliptic integral of the second kind: $E(m)\equiv E(K(m)|m)$. Using these two facts, we find that 
\begin{equation}
\int_{\rm period} W =(2-m-a) K(m) - 2 E(m)
\end{equation}
Thus, choosing 
\begin{equation}
a=2-m-2E(m)/K(m)
\label{special}
\end{equation}
leads to unbroken SUSY. With this choice for the parameter $a$, both $H_\pm$ have periodic Bloch zero-modes, 
\begin{equation}
\psi_0^{(\pm)}=exp\left\{\pm\left[m sn(x|m) cd(x|m)- 2Z(x|m)\right]\right\}
\label{ellzero}
\end{equation}
where $Z(x|m)$ is the Jacobi zeta function, which is related to the elliptic integral $E(x|m)$ through $E(x|m)=Z(x|m)+x E(m)/K(m)$. 

Figure 1 contains plots of both $W$ and $W^2$, showing that the tree-level potential $W^2$ has two zeros within each period. These plots have been made for the particular choice (\ref{special}) for the parameter $a$. However, these plots are representative. In fact, if $0<a<m^2/(1+\sqrt{1-m})^2$, $W^2$ always has two zeros within each period. The choice (\ref{special}) falls within this range. The corresponding bosonic and fermionic potentials (\ref{iso}) are plotted in Figure 2. Notice that the two potentials are simply parity reflections of one another - they are ``self-isospectral'' in the sense of \cite{dunne}. The zero-mode wavefunctions (\ref{ellzero}) are plotted in Figure 3. Notice that they are smooth, bounded, periodic, and have no zeros. 

This model has an interesting $m\to 1$ limit. (Note that the period $K(m)$ diverges logarithmically \cite{abram} as $m\to 1$.) With the parameter choice in (\ref{special}), SUSY is unbroken for any $m<1$. However, when $m=1$, $W=-sech^2x$, which is an even function on the real line, so that SUSY is broken. Indeed, the zero modes in (\ref{ellzero}) become $\psi_0^{(\pm)}=exp\left\{\mp tanh\, x\right\}$, neither of which is normalizable on the real line. This type of discontinuous behavior is interesting because we could alternatively (as was done for the model (\ref{ell-1}) in \cite{braden}) consider this not as a periodic potential model, but as a model defined on a finite interval with periodic boundary conditions, which would therefore have a discrete (rather than a band) spectrum. This type of regularization in a finite spatial volume is a common computational device. The conventional wisdom \cite{witten} is that if SUSY is shown to be unbroken in any finite volume, then this will persist in the infinite volume limit. Here, however, SUSY becomes broken in the infinite volume limit (i.e. when $m=1$), even though it is unbroken for any finite volume (i.e. for any $m<1$). How can this be happening? The answer is that the zeros of the tree-level potential $W^2$ within a single period, $-\frac{K(m)}{2}<x\leq \frac{K(m)}{2}$, (see Figure 1) disappear to infinity in the infinite volume limit (i.e. as $m\to 1$). Indeed, when $m=1$, $W^2=sech^4x$, which has no zeros at all. Thus, these vacua are receding to infinity and no longer play any role in the Hilbert space of the infinite volume theory. It is interesting that this type of behavior, with vacua disappearing to infinity, appears in the massless limit of SUSY QCD \cite{davis}.

To conclude, we have shown that SUSY quantum mechanics models with periodic superpotentials are more sensitive to tuning of the parameters than are the more familiar nonperiodic models. Since the generic insensitivity of the Witten index to fine-tuning of parameters is often invoked in investigations of SUSY breaking in field theories, it would be interesting to learn whether these periodic models have field theoretic analogues.

\vskip .5in
\noindent{\bf Acknowledgements}
This work has been supported by the DOE grant DE-FG02-92ER40716.00 (GD), and by the NSF (JM) through the R.E.U. (Research Experience for Undergraduates) Grant PHY-9424125 (W. Stwalley, P.I.) at the University of Connecticut.

\begin{figure}
    \epsffile{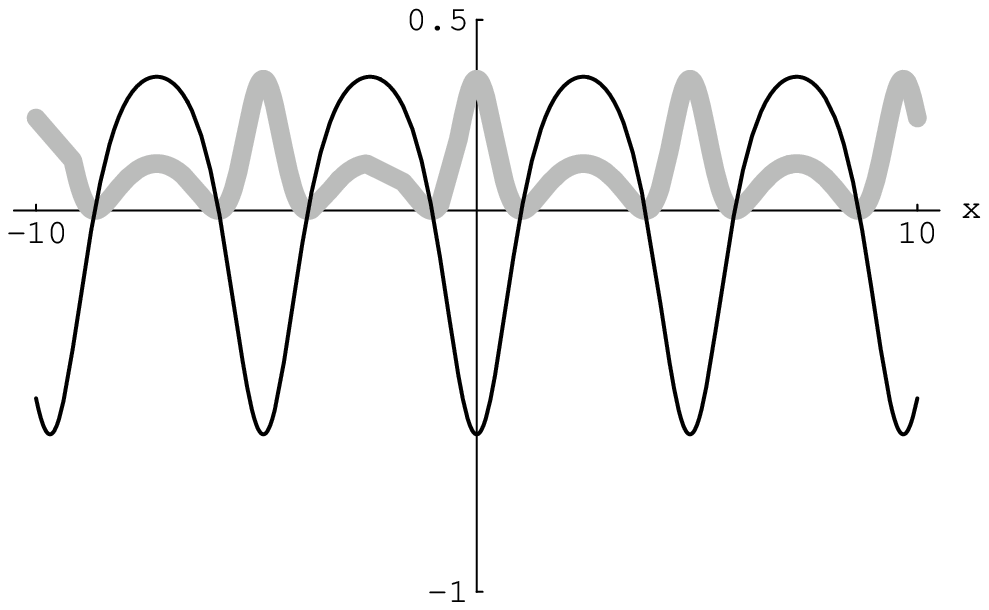}
    \caption{The superpotential W(x) (thin line) in (\protect{\ref{ell-2}}), and its square $W^2(x)$ (thick line).These plots are for the elliptic parameter m=.999, for which the period is K(m)=4.84. Notice that $W^2$, which is the tree-level potential, has two zeros within each period.}
  \label{plot1}
\end{figure}

\begin{figure}
    \epsffile{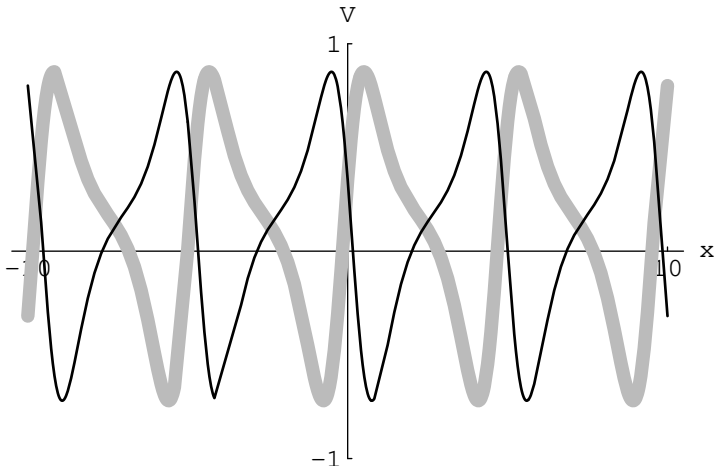}
    \caption{The bosonic potential $V_+$ (thick line) and the fermionic potential $V_-$ (thin line) for the system with superpotential in (\protect{\ref{ell-2}}), with elliptic parameter m=.999. Notice that the two potentials are simply parity reflections of one another - they are ``self-isospectral'' in the sense of \protect{\cite{dunne}}.}
  \label{plot2}
\end{figure}

\begin{figure}
    \epsffile{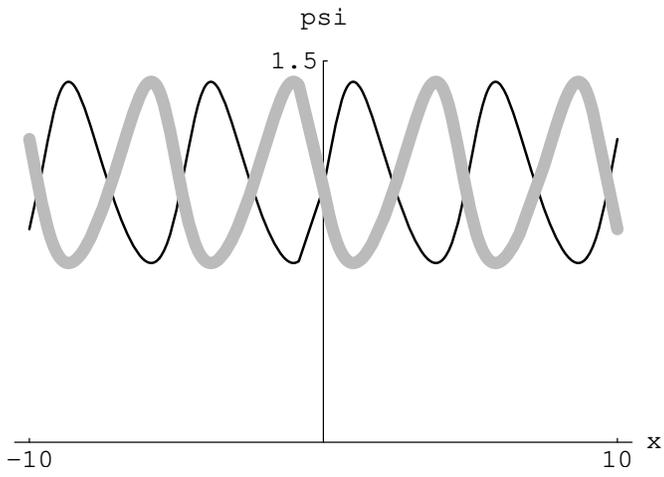}
    \caption{The zero-modes $\psi_0^{(+)}$ (thick line) and $\psi_0^{(-)}$ (thin line) in (\protect{\ref{ellzero}}). Notice that they are smooth, bounded, periodic, and have no zeros.}
  \label{plot3}
\end{figure}

\end{document}